\documentclass[aps,pre,twocolumn,floatfix]{revtex4}
\usepackage{graphicx, amsmath}
\usepackage{amsfonts}
\usepackage{color}

\begin{document}

\title{The phase diagram of random threshold networks}
\author{Agnes Szejka, Tamara Mihaljev and Barbara Drossel}
\address{Institut f\"ur Festk\"orperphysik, TU Darmstadt, Hochschulstrasse 6, 64289 Darmstadt, Germany}

\begin{abstract}
  Threshold networks are used as models for neural or gene regulatory
  networks. They show a rich dynamical behaviour with a transition
  between a frozen and a chaotic phase. We investigate the phase
  diagram of randomly connected threshold networks with real-valued
  thresholds $h$ and a fixed number of inputs per node. The nodes are
  updated according to the same rules as in a model of the cell-cycle
  network of \textit{Saccharomyces cereviseae} [PNAS \textbf{101},
  4781 (2004)]. Using the annealed approximation, we derive
  expressions for the time evolution of the proportion of nodes in the
  ``on'' and ``off'' state, and for the sensitivity $\lambda$. The
  results are compared with simulations of quenched networks. We find
  that for integer values of $h$ the simulations show marked
  deviations from the annealed approximation even for large networks. This can be attributed
  to the particular choice of the updating rule.
\end{abstract}
\maketitle

\section{Introduction}
Threshold networks can be used to model gene regulatory networks
\cite{Bornholdt00, Li04, Sevim08}. The nodes of the network represent
genes, and the directed links between them represent interactions
between genes. Each node $i$ can be in two different states $\sigma_i=
1, 0$ (``on'', ``off'').  That means that the gene is either expressed
or not expressed.  Furthermore, each node receives inputs from $K$
randomly chosen other nodes that regulate its activity cooperatively.
The interactions between the nodes can be excitatory or inhibitory so
that one node can activate or repress the expression of another node.
In this paper, we study a threshold network where the time development of the states of the nodes is given by the
following equation
\begin{eqnarray}
\sigma_i(t+1) =
\left \{ \begin {array}{ll}
1, & \sum \limits_j c_{ij}\sigma_j(t) - h > 0 \\ 
0, & \sum \limits_j c_{ij}\sigma_j(t) - h < 0 \\ 
\sigma_i(t), & \sum \limits_j c_{ij}\sigma_j(t) - h = 0\, .
\end{array} \right. \label{eins}
\end{eqnarray}
Here, $h$ is a threshold that is the same for every node. The
couplings $c_{ij}$ are $\pm 1$ with equal probability, $c_{ij} = 0$ if
node $i$ receives no input from node $j$. The input $s_{ij} =
c_{ij}\sigma_j$ from node $j$ to node $i$ can therefore take three
different values: $0, +1$ or $-1$. A node becomes activated when the
sum of its inputs exceeds the threshold value, and it becomes inactive
when the sum of its inputs is below the threshold. When the sum of the
inputs gives exactly the threshold value the node does not change its
state in the next time step. The nodes are updated in parallel. These
dynamics with $h = 0$ and a $K$ value that varies from node to node
were used to model the cell-cycle network of \textit{Saccharomyces
  cereviseae}.  This model was able to reproduce the overall dynamic
properties of the real network \cite{Li04}. There exist several
variants of threshold models. In other variants, the $c_{ij}$ can be
continuous quantities chosen at random from some probability
distribution; the spin values may be $\pm 1$ instead of 1 and 0 (see
for instance \cite{Kuerten88b, Rohlf02, Rohlf07}), or the update rule
in the case that the sum of the inputs is exactly at the threshold can
be different.  

Models that use spin values $\sigma_i = 1, 0$  can be mapped onto models with spin values $r_i= \pm 1$ 
 by making the substitution  $\sigma_i = (r_i+1)/2$. For our update rule (\ref{eins}), this leads to
\begin{eqnarray}
r_i(t+1) =
\left \{ \begin {array}{ll}
1, & \sum \limits_j c_{ij}r_j(t) >2h - \sum \limits_j c_{ij} \\ 
-1, & \sum \limits_j c_{ij}r_j(t)<2h -\sum \limits_j c_{ij}  \\ 
r_i(t), & \sum \limits_j c_{ij}r_j(t)=2h-\sum\limits_j c_{ij} 
\end{array} \right.
\end{eqnarray}
This means that each node $i$ obtains its own threshold value $h_i$,
which depends on the values of the $c_{ij}$.  Therefore the dynamics
of the model studied in this paper is different from that of the $\pm
1$ model studied more widely.

Similarly to random Boolean networks, random threshold networks show a
transition between a frozen and a chaotic phase when the network
parameters are varied. In the frozen phase, a perturbation at one node
propagates during one time step on an average to less than one other
node. In the chaotic phase, the difference between two initially
almost identical states increases exponentially fast, because a
perturbation propagates on an average to more than one node during one
time step. In the frozen phase, the length of attractors (i.e., the
number of states on attractors) is either 1 or very small. Most of the
nodes are frozen, that is they do not change their states anymore in
the stationary state. In the chaotic phase, attractors are very long
on average, and a non-vanishing proportion of the nodes change their
states on the attractors.  This phase transition was previously
studied in threshold networks in \cite{Kuerten88b, Rohlf02, Rohlf07}.

\section{The phase diagram}
With the help of the annealed approximation introduced by Derrida and
Pomeau \cite{Derrida86}, one can determine the parameter values $h$
and $K$ for which the networks are in the chaotic or in the frozen
regime. This approximation neglects that the input connections to
nodes are constant in time (quenched). It describes therefore a situation where the connections are changed randomly in
each time step. The annealed approximation also neglects fluctuations and
can therefore become exact only for infinitely large networks (if at
all). The parameter $\lambda$, called sensitivity \cite{Luque97,Shmulevich04},
which is $K$ times the average probability that the output of a node changes
when one of its inputs changes, discriminates between the two
phases. If $\lambda < 1$, the network ensemble is
said to be in the frozen phase. If $\lambda > 1$ it is in the chaotic
phase. For $\lambda = 1$ the networks are critical. In order to determine
$\lambda$, one has to know $b_t$, the proportion of nodes in state 1 at
the considered moment in time. $\lambda$ is a function of $b_t$ and
becomes constant only when $b_t$ has reached a fixed point.

The annealed approximation has been used successfully to predict the
phase diagram of various classes of random Boolean networks. In those
networks, correlations between nodes are apparently irrelevant for the
evaluation of $b_t$ and $\lambda$. We will see further below that this
is not correct for all threshold networks. 

\subsection{Time evolution of $b_t$}
Let us first calculate $b_{t+1}$ as function of $b_t$ using the annealed approximation.
For non-integer $h$, the value of a node will be 1 in the next time step if the sum of its inputs is larger than $h$. Therefore, 
\begin{equation}
\begin{split}
b_{t+1}^{(1)} = \sum \limits _{m = \lfloor h \rfloor +1}^K {K \choose m} \left[ \sum \limits _{l = \lfloor{\frac{m+h}{2}}\rfloor + 1} ^{m} {m \choose l} \right] \cdot \\
\cdot \left(\frac{b_t}{2}\right)^m \left( 1 - b_t\right)^{K-m}.
\label{bt}
\end{split}
\end{equation}
Here, $m$ is the number of input nodes with value 1. For positive threshold values at least $\lfloor
h \rfloor + 1$ of the input nodes must be active if the sum of the inputs shall
be larger than the threshold. For the same reason, the number $l$ of positive (excitatory)
couplings from these active input nodes must be at least
$\lfloor{\frac{m+h}{2}}\rfloor + 1$. For negative $h$ values all configurations with $m < |h|$ also contribute to the sum. There are ${K \choose m}$
different possibilities to choose $m$ active nodes among the $K$ input
nodes and ${m\choose l}$ different possibilities to choose $l$ 
excitatory links among the links from these active nodes. Finally
${b_t}^m \left( 1 - b_t\right)^{K-m}$ is the probability that $m$
input nodes are in state $\sigma_j = 1$ and the others in state $\sigma_j
= 0$, and $\left( \frac{1}{2} \right)^m$ is the probability that
positive and negative couplings are distributed as they are.

For integer-valued $h$, the sum of the inputs can be exactly at the
threshold, which is not possible for non-integer $h$. Within the annealed approximation, a node with the inputs at the threshold 
will be ``on'' with a probability $b_t$ in the next time step. Equation (\ref{bt}) obtains therefore a second term when $h$ is integer and becomes
\begin{equation}
\begin{split}
b_{t+1}^{(2)} = b_{t+1}^{(1)} + \sum \limits _{l = h} ^{\lfloor \frac{K + h}{2} \rfloor} {K \choose {2l - h}} {{2l - h} \choose l} \cdot \\
\cdot \left( 1 - b_t \right)^{K - 2l + h} \left( \frac{b_t}{2} \right)^{2l - h} \cdot b_t.
\label{btint}
\end{split}
\end{equation}
Here, $l$ is again the number of active input nodes with positive
couplings. The number of active nodes with negative couplings has to
be $l - h$ in order to place the sum of the inputs at the threshold.

\begin{figure}
\hspace{-0.7cm}\includegraphics[angle = -90, width = \columnwidth]{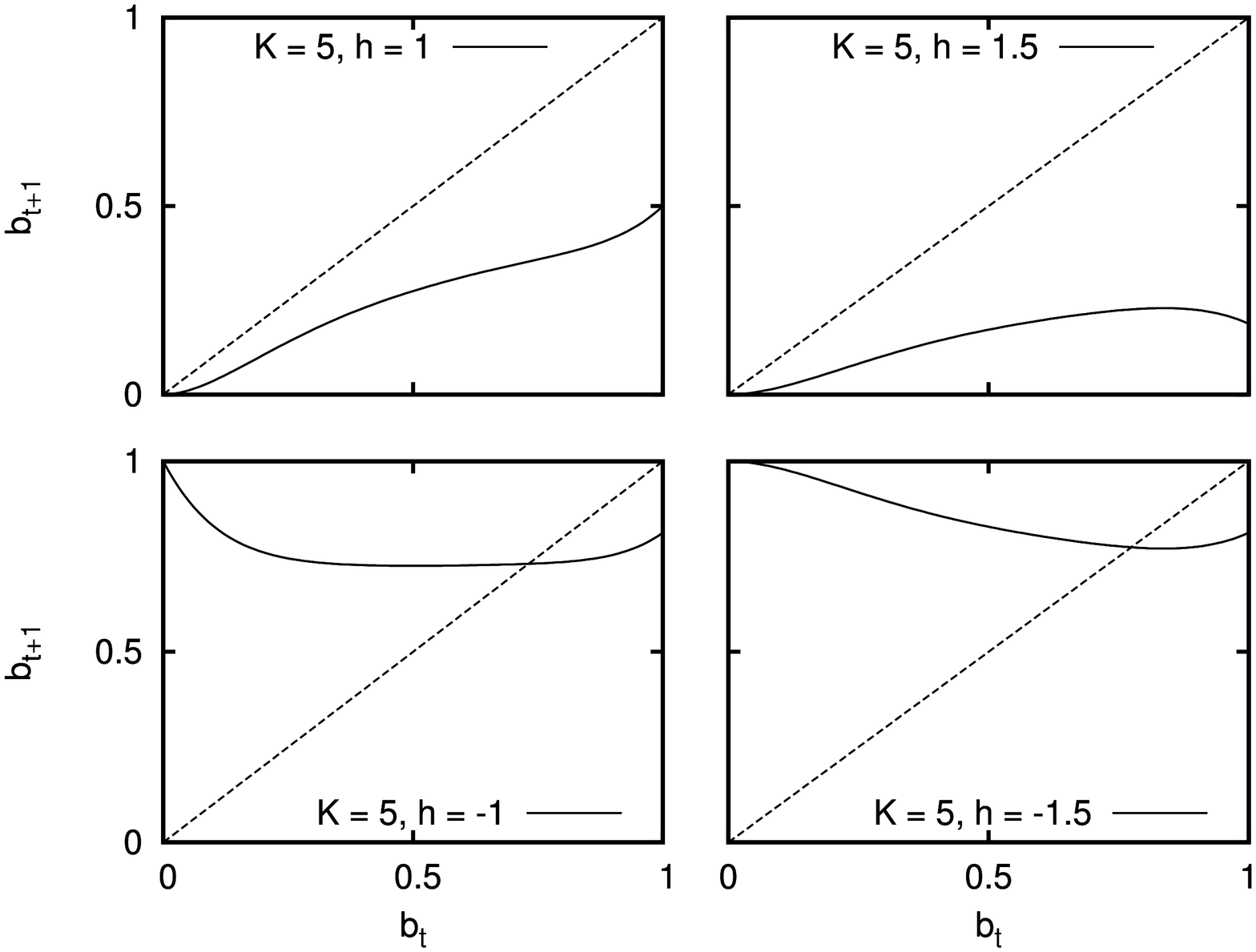}
\caption{Maps $b_{t+1}$ versus $b_t$ for $K = 5$ and different $h$. The dashed line is the bisector.}
\label{btmap}
\end{figure}

Having established the recursion relation for $b_t$, one can plot maps
$b_{t+1}$ vs. $b_t$ for different $h$ and $K$. The fixed points
$b^*$ of the map (see figure \ref{btmap}) are stationary
solutions of the annealed approximation.

A fixed point $b^*=0$ exists whenever the smallest $m$ contributing to
the sum in (\ref{bt}) is larger than 0. This is the case for all $h
\ge 0$. The fixed point $b^*=0$ is stable when the slope of the map at $b_t=0$
is smaller than 1, which is the case for all $h \ge 1$. For $0<h<1$,
the map is $b_{t+1}=Kb_t/2$ to leading order in $b_t$, and the fixed
point $b^*=0$ is therefore unstable for $K>2$. For $K=2$, we have to
include the next order in $b_t$, which gives $b_{t+1}=b_t(1-3b_t/4)$,
and therefore the fixed point $b^*=0$ is stable. For $h=0$, we have to
leading order $b_{t+1}=b_t(1+K/2)$, and the fixed point $b^*=0$ is
therefore unstable.

\begin{figure}
\hspace{-0.7cm}\includegraphics[angle = -90, width = \columnwidth]{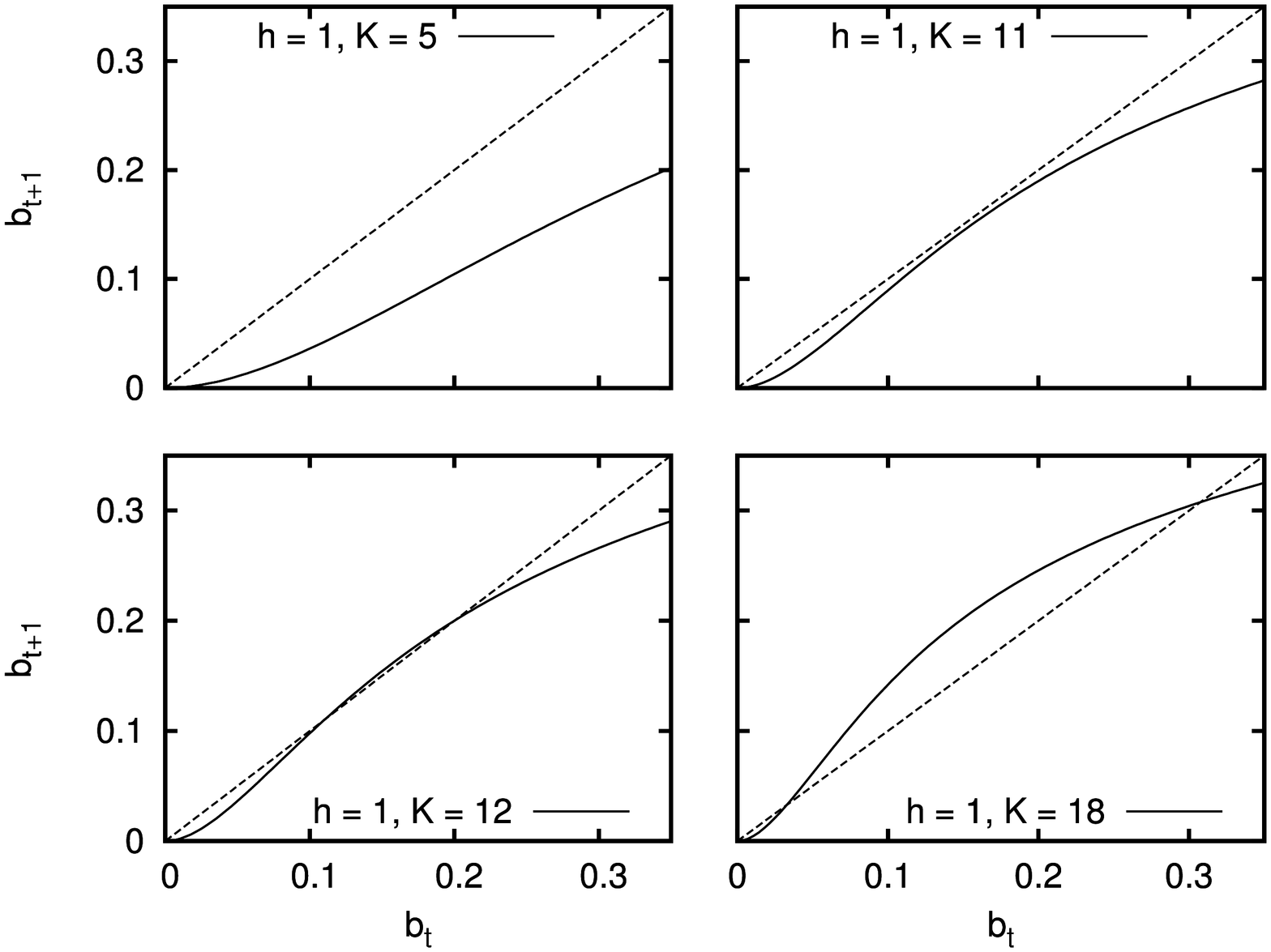}
\caption{Maps $b_{t+1}$ versus $b_t$ for $h = 1$ and different $K$. The dashed line is the bisector.}
\label{maph1}
\end{figure}

In order to obtain information about other fixed points, we iterated
numerically the recursion relations for $b_t$ and plotted the map. We
found that for $h \geq 1$ and sufficiently small $K$, the only stable
fixed point is $b^{*(1)} = 0$, but for growing $K$ a second stable
fixed point appears. Figure \ref{maph1} shows how the map changes with
increasing $K$ when $h=1$. A second stable fixed point with $b^{*(2)}
= 0.200$ appears at $K = 12$. It moves with increasing $K$ slowly
towards the value $0.5$, which is the asymptotic value for
$K\to\infty$.  For $0<h <1$ and $K = 2$, the only fixed point is
$b^{*} = 0$.  For $K > 2$, this fixed point is unstable, as mentioned
before, and there exists a stable fixed point with a value $0 < b^*
\leq 0.5$. It moves towards $0.5$ with increasing $K$. For $h = 0$,
there is only one stable fixed point $b^{*} = 0.5$ for all values of
$K$.  For $h < 0$, the stable fixed point lies between 0.5 and 1 and
moves towards 0.5 with increasing $K$.

For $K=1$, all fixed points can be determined analytically. For $h>1$,
the input of no node can be above the threshold, and there is no fixed
point besides $b^{*} = 0$. Similarly, for $h<-1$ the inputs of all
nodes are above the threshold, and therefore $b^* =1$. Evaluation of
the recursion relation for the remaining values of $h$ gives 
$b^{*} = 1$ for $h=-1$ and $b^*=2/3$ for $-1<h<0$ and $b^*=1/2$ for
$h=0$ and $b^*=0$ for $h>0$. 

\subsection{$\lambda$}
Having found a fixed point $b^*$ for a  pair of parameter values $h$
and $K$, one can determine the corresponding value of $\lambda $. For
non-integer $h$, the change of an input can affect the output only when the other $K-1$ inputs sum up to be directly above or directly underneath the threshold. This leads to
\begin{equation}
\begin{split}
\lambda = \frac{K}{2} \cdot \sum \limits_{m = \lfloor |h| \rfloor}^{K-1} {{K-1}\choose m} {m \choose {\lfloor \frac{m+1+h}{2} \rfloor}} \cdot \\
\cdot \left( 1-b^* \right)^{K-1-m} \left( \frac{b^*}{2} \right)^m
\end{split}
\end{equation}
Here, $m$ is again the number of active input nodes. The number of
active input nodes with positive couplings, $l = \lfloor
\frac{m+1+h}{2} \rfloor$, is chosen in such a way that $2l-m$ is close to
the threshold (directly underneath or directly above). The factor
$1/2$ is the probability that the $K$th coupling has the proper sign.

For integer $h$, the situation is again different. A change in an
input can affect the output only if the sum of all inputs was $h$
before the change. In the opposite situation, where a change in an
input places the total input exactly at the threshold, the output does
not change. (If we took the annealed approximation to its extremes and
ignored the fact that there is a correlation between the state of a
node and the state of its inputs, we would need to consider also the
case that the sum of the inputs is $h \pm 1$ before the change of one
input node.)

We therefore obtain for integer $h$
\begin{equation}
\begin{split}
\lambda = \frac{K}{2} \cdot  \sum \limits_{l = h} ^{\lfloor
    \frac{K+h}{2} \rfloor} {K \choose {2l-h}} {{2l-h} \choose l} \cdot \\
\cdot \left(1-b^*\right)^{K-2l+h} 
\left( \frac{b^*}{2}
  \right)^{2l-h} 
\label{hint}
\end{split}
\end{equation}

\begin{figure}
\hspace{-0.7cm}\includegraphics[width = \columnwidth]{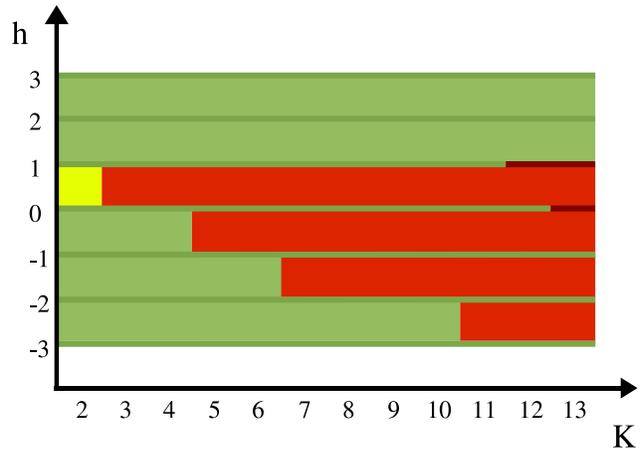}
\caption{Phase diagram obtained from the annealed approximation. The chaotic phase ($\lambda > 1$) is indicated in red, the frozen phase ($\lambda < 1$) in green, and the critical region ($\lambda = 1$) in yellow. For integer $h$ the phases are indicated by darker shades of the colours. (In black and white the lighter colours indicate the frozen phase, the darker colours the chaotic phase.)}
\label{phase}
\end{figure}

Using the last two equations, one can evaluate $\lambda$ for every
combination of $h$ and $K$. The resulting phase diagram is shown in
Figure \ref{phase}. Only networks
with $K = 2$ and a threshold value $0 < h < 1$ are critical. Networks
with $h > 1$ are frozen in the $K$ range shown. Where there are two
stable fixed points for $h \geq 1$ networks are frozen at $b^* = 0$
and chaotic at $b^* > 0$. For $h < 1$, networks with integer $h$ are
more ordered than those with non-integer $h$.

\section{Numerical Simulations}
We performed computer simulations of quenched threshold networks for
different $h$ and $K$ and compared the results to those obtained in
the framework of the annealed approximation.  The number of nodes was
$N = 10^5$ in all simulations. The threshold values were chosen in
the range $-3 < h < +3$, with non-integer thresholds chosen to be $h =
-2.5, -1.5, \dots, 1.5, 2.5$. All non-integer thresholds that lie
between the same two integers lead obviously to the same dynamical
behaviour, therefore it is sufficient to consider these values.

We will first look at the value $b_f$ found after a sufficiently long
time when starting with some initial proportion $b_0$ of 1s in the
system.

\subsection{The proportion of 0s and 1s}
\subsubsection{$h > 0$}
For non-integer $h$, the simulations are in good agreement with the
predictions of the annealed approximation.

$h = 0.5$: The only fixed point for $K = 2$, $b^* = 0$, is
 weakly stable because the map has a slope of 1 at this point.
Most simulated networks do not reach this fixed point but run into
attractors of varying length with a small $b_f$ of the order of
$10^{-2}$. This is due to the fact that the iteration formula
$b_{t+1}=b_t(1-3b_t/4)$ can be applied only as long as $b_t^2$ is
larger than of the order $1/N$. The negative quadratic term in this
equation describes the repressive effect of a second active input with
a negative coupling. The decrease of $b_t$ comes to a halt when $b_t$
has become so small that there are no more nodes with two active
inputs, which will happen for ever smaller values of $b_t$ when the
system size is made larger. The discrepancy between the simulations and
the annealed approximation is thus clearly a finite-size effect.

For $K = 3$ and $4$, the mean $b$ values of the quenched networks are in
good agreement with the calculated values (we checked for agreement in
three decimal places).

$h = 1.5$: For non-integer $h$, the nonzero fixed point value $b^*$
appears at $K$ values that are in accordance with the annealed
approximation. For $h = 1.5$, this happens at $K = 15$. The average
$b$ value obtained from our simulations shows however a small
deviation of about 1\% from the calculated value. We can ascribe this
small discrepancy again to finite-size effects, since the slope of the
map near the fixed point is close to 1 for the $K$ value where this
fixed point occurs first. For $K = 16$ and $17$, the values are again in
good agreement.

$h = 2.5$: For $h = 2.5$, a stable fixed point value $b^*>0$ appears
at $K = 40$, just as predicted by the calculations. The value of $b^*$
obtained from the simulations and averaged over the attractor agrees
with the one obtained from the annealed approximation in three decimal
places. Such a good agreement is also found for $K = 41$ and 42.

The case of \textit{integer $h$} is special, and we will see that in
this situation the simulations are not in good agreement with the annealed
approximation.

\begin{figure}
\hspace{-0.7cm}\includegraphics[angle = -90, width = \columnwidth]{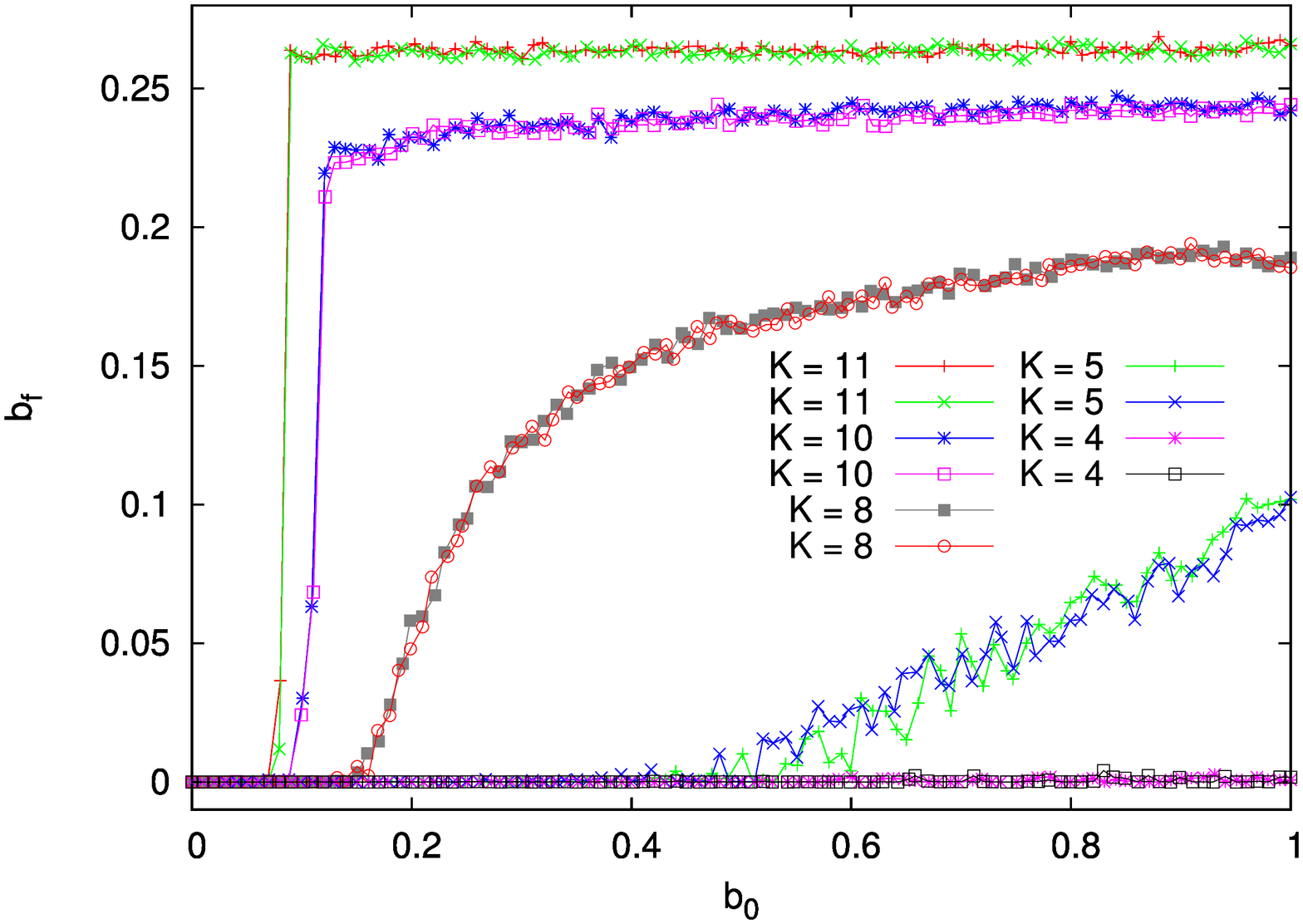}
\caption{The final proportion of 1s in dependence of the initial proportion of 1s in networks with $h = 1$ and different values of $K$. Each data set corresponds to one network realization.}
\label{h1}
\end{figure}

$h = 1$: For networks with a $K$ value ranging from 2 to 11, the
annealed approximation predicts a single stable fixed point $b^* = 0$.
In contrast, our simulations show that already for $K = 5$ there exist
stationary states with a larger number of active nodes, and the value
$b^*$ approached for large times depends on the initial value $b_0$
(see figure \ref{h1}). Each curve in figure~\ref{h1} corresponds to one
network realization. Each point in the figure is a fixed point of the
dynamics, that is an attractor of length one.  As one can see, two
different networks with the same values of $K$ and $h$ show
approximately the same behaviour. This means that the function
$b_f(b_0)$ does not depend on the detailed realization of the
networks. For networks with $K = 5$,  the nonzero fixed point appears when 
the initial proportion of 1s is around 50\%.

Only for $K \ge 12$, the annealed approximation predicts a stable
fixed point $b^*>0$. For such $K$ values, our simulations give a value
$b_f$ that is independent of $b_0$ if $b_0$ is not too small. If one
compares the value $b^*$ obtained by iterating formula \ref{btint}
with the value obtained by averaging over several simulated networks,
one finds that the quenched networks have around 41\% more active
nodes than predicted by the annealed approximation. in the
publications cited above, where all nodes have the same threshold
value. With increasing $K$, this deviation from the annealed
approximation decreases. It is about 25\% for $K = 13$ and about 20\%
for $K = 14$.

In order to understand how a broad set of dynamical fixed points can
emerge in these systems, we note the following: (a) At every fixed
point, there are nodes the sum of whose inputs is at the threshold.
If changing the state of such a node does not change the state of any
node influenced by it, we have found another fixed point with a
different number of active nodes. (b) All fixed points of a network
with $h=1.5$ or with $h=0.5$ are also fixed points of this network if
$h=1$. One can expect that all observed fixed points should have $b$
values between these two boundary values, and this is indeed observed.
(c) If there exists a set of fixed points with different $b$ values,
the final value of $b$ reached in a simulation should depend on the
initial value. This in turn means that the state of a node at a fixed
point depends on the dynamical history of this node and its inputs.
Such correlations between the states of a node at different moments in
time are not taken into account in an annealed approximation, which is
therefore no good approximation in this situation. With increasing
$K$, the curves in figure \ref{h1} become increasingly independent of
$b_0$.  This can be attributed to increasing transient times, which
weaken the ``memory'' of the initial state of the nodes. (d) There can
exist sets of active nodes that are connected by positive directed
links among each other in a way that loops are formed. When the sums
of all other inputs to the nodes in such loops are zero, they are at
the threshold and once activated will keep their value. Correlations
of this type between nodes are not included in annealed models.

$h = 2$ and $3$: For networks with a threshold value $h = 2$, a
second stable fixed point value $b^*$ appears at $K = 35$ in the annealed
approximation, but can be found in quenched networks already for $K =
31$. At $K = 35$, the second stable
fixed point $b^*$ is about 36\% higher than the value predicted by the annealed approximation, and this difference decreases to 26\% and 21\%  for $K = 36$ and $K = 37$. For $h = 3$, a second stable fixed point
appears also at a lower $K$ (at $K=65$) than expected from the annealed
approximation, where it appears at $K=70$. The value of $b^*$ reached
at $K = 70$ is about 29\% higher than the expected value. We find a
deviation of about 23\%  and 20\% for $K = 71$ and $K = 72$.
All observed $b$ values are between those obtained
for $h=1.5$ and $h=2.5$ if $h=2$, and between those obtained
for $h=2.5$ and $h=3.5$ if $h=3$.

\subsubsection{$h = 0$}
\begin{figure}
\hspace{-0.7cm}\includegraphics[angle = -90, width = \columnwidth]{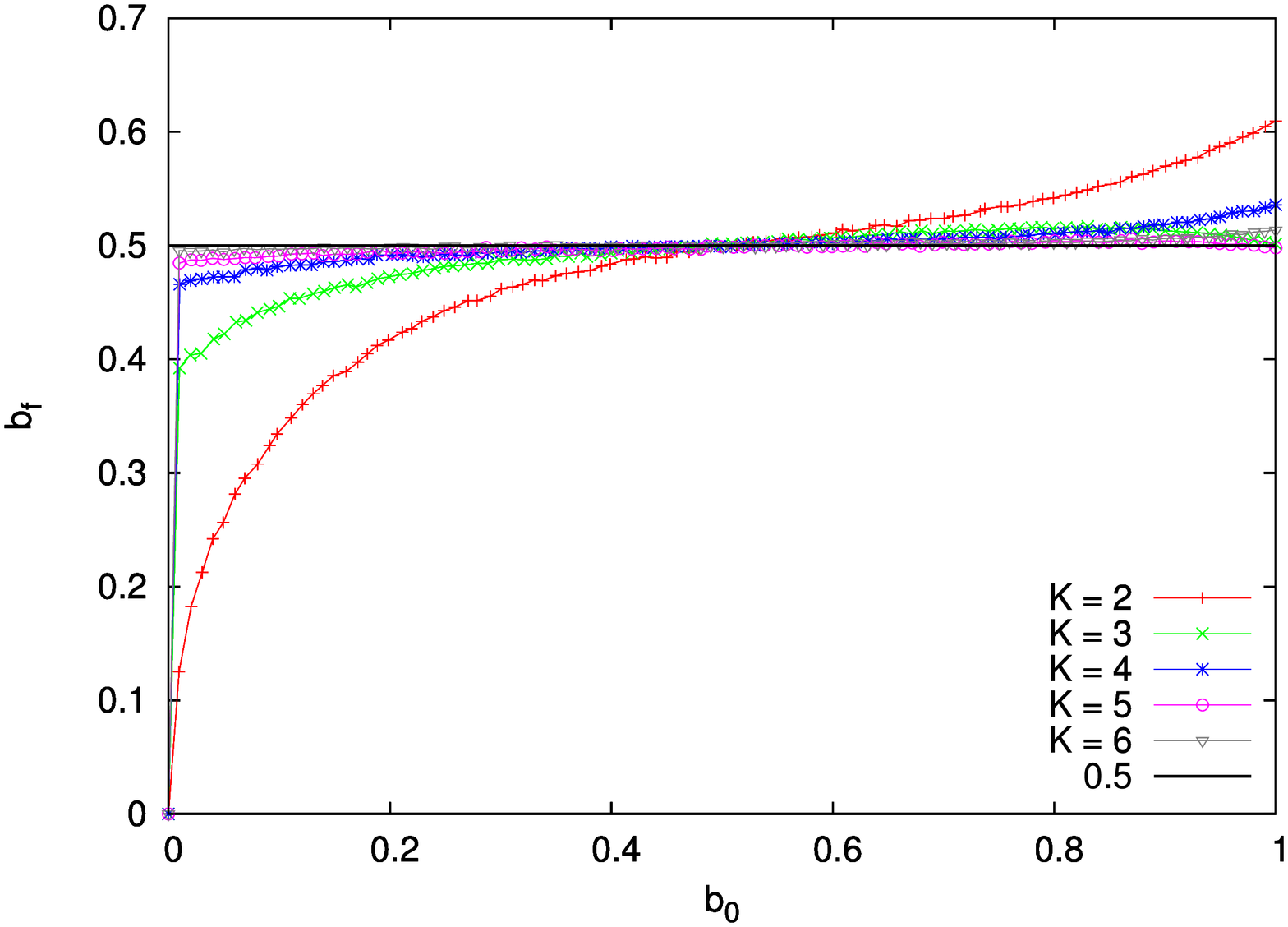}
\caption{The final proportion of 1s in dependence of the initial
  proportion of 1s in networks with $h = 0$ and different values of $K$. 
Each data set corresponds to one
network realization.}
\label{h0}
\end{figure}
For $h = 0$ and $K$ ranging from $K = 2$ to 5 we again find many
dynamical fixed points that have a $b_f$ different from the value
calculated with the help of the annealed approximation, see figure \ref{h0}.
With increasing $K$ however, $b_f$ approaches the expected
value. 

In Figure \ref{h0}, one can observe another interesting effect, which
occurs also for all other integer $h$ and for not too large $K$
values.  Depending on the sum of $h$ and $K$, the curves exhibit
different behaviour at high $b_0$ values. If $(h+K)$ is even, $b_f$
increases at the end as in this case it is more likely that the inputs
of a node are at the threshold than for odd $(h+K)$.

Considering the trivial case $K=1$ is also instructive: 
Every node has exactly one input. Starting from a randomly chosen
node, we can follow the chain of inputs preceding this node. 
For large system sizes, the average length of such chains is long (it is
of the order of $\sqrt{N}$)~\cite{flyvbjerg:exact}. Every chain eventually ends in a loop.
Along the chain, positive and negative couplings follow in a random
order. We can easily find the fixed points of such a system of chains: 
Obviously, having all nodes in state 0 is a fixed point of the
dynamics. If we then switch on a node that has only negative output
links, we obtain another fixed point. If we switch on a node that has
a positive output link, the node influenced through this link must
also be switched on, and so on, until the end of the sequence of
positive couplings is reached. The fixed point with the maximum number
of ``on'' nodes is obtained by 
assigning a 1 to all nodes with a positive input link and
to all those nodes with a negative input link that are preceded by an
odd number of nodes with a negative input link. In this way, all nodes
that have a negative input link and an input node in state 1, are
in state 0. The $b$ value associated with this fixed point is
$$ b\simeq \frac 1 2 + \frac 1 8 \sum_{n=0}^\infty \left(\frac 1 4
\right)^n = \frac 2 3 \ . $$ If we had $-1 < h < 0$, this would be the
only fixed point. The example $K=1$ thus demonstrates that the maximum
possible fixed point value for $b$ for integer $h$ is identical to the
one obtained by slightly lowering the value of $h$. Similarly, the
minimum value 0 is the fixed point value obtained if $h$ is slightly
larger than 0.  All intermediate values of $b$ for $h=0$ are obtained by
switching off part of the ``on'' nodes in the state with the maximum
$b$ value.

This example demonstrates also that not all fixed point values of
$b$ can be reached from a random initial state. For instance, the maximum
value 2/3 of $b$ cannot be reached by starting from such a random
initial configuration. If initially all nodes are in state 1, i.e. if
$b_0=1$, we have after 1 time step $b=1/2$, where all nodes with a
positive input link are in state 1. If there are less 1s initially,
there cannot be more 1s in the final state.  Values between $1/2$ and
$2/3$ can therefore only be reached by starting from specially
prepared initial states. This observation for $K=1$ explains why the
simulations for larger $K$ (and also for other integer $h$) give $b$
values between those obtained with the annealed approximation for the
neighbouring non-integer $h$ values, but not the entire $b$ interval
between these boundaries is reached from a random initial state with
fixed $b$.

The fact that the agreement between the annealed approximation and the
quenched networks becomes better for larger $K$ can be ascribed to the
narrowing of the interval between the boundaries with growing $K$.

\subsubsection{$h < 0$}
\begin{figure}
\hspace{-0.7cm}\includegraphics[angle = -90, width = \columnwidth]{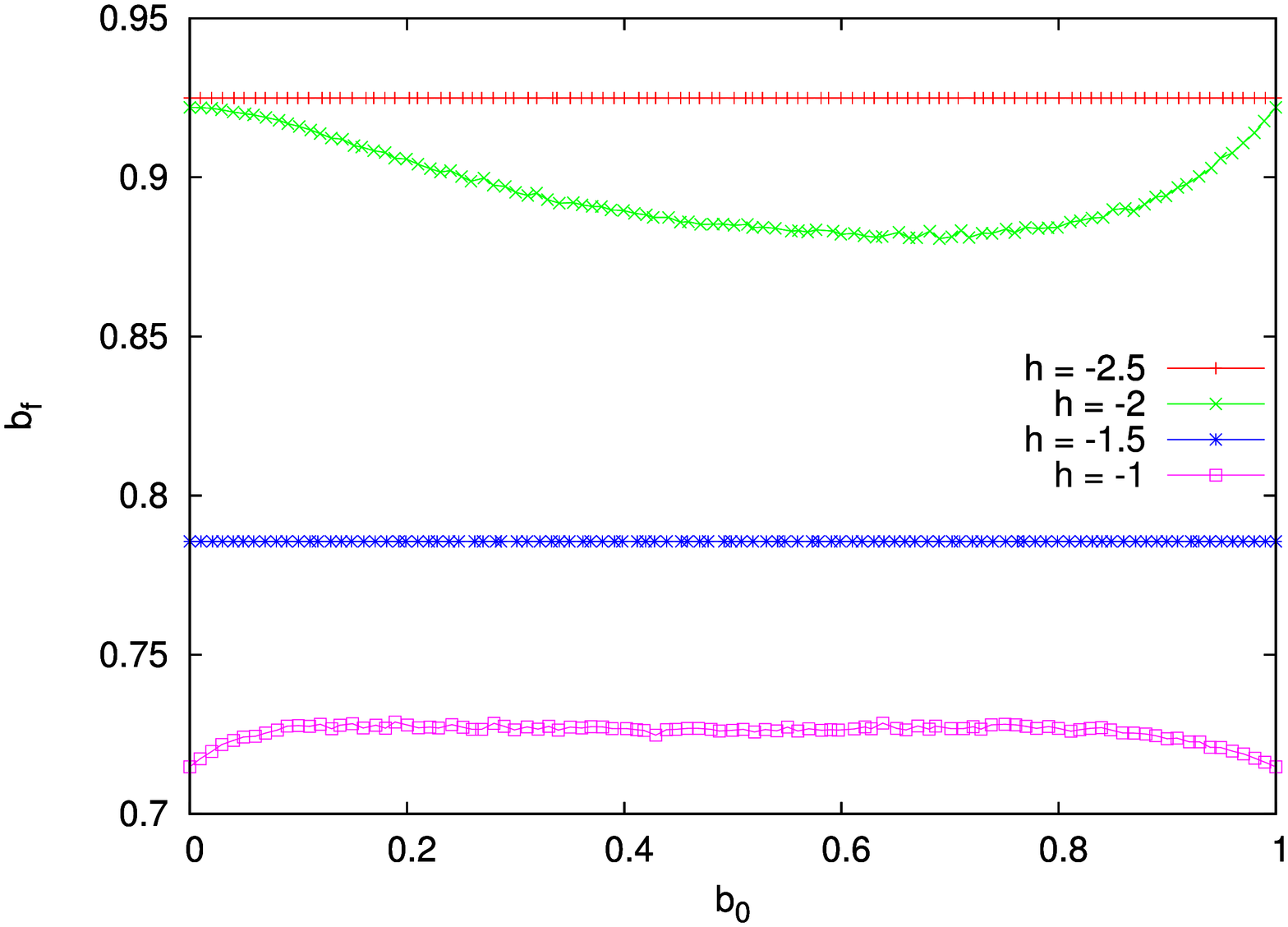}
\caption{The final proportion of 1s in dependence of the initial proportion of 1s in networks with $K = 4$ and different negative $h$. Each data set corresponds to one network realization}
\label{hneg}
\end{figure}
Figure \ref{hneg} shows the final $b$ value as a function of the initial
$b$ value for networks with $K = 4$ and different negative $h$. (For
the parameter values shown all networks have $\lambda < 0$, that is
they are frozen.) As one can see, for non-integer $h$ the final
proportion of 1s in the network does not depend on $b_0$. Again the
values of $b_f$ obtained with the simulations agree well with those
obtained using the annealed approximation.
For integer $h$, we find dynamical fixed
points with a value  $b_f$ that depends on $b_0$. The mean value of $b_f$
is always smaller than the value predicted by the annealed
approximation. The values of $b_f$ for the networks with integer $h$
always lie between those of networks that have
neighbouring non-integer $h$ values with the same $K$, as we have also
observed for non-negative integer values of $h$.

\subsection{The phase diagram}
Next, we report on the dynamical properties of the simulated networks,
that is the lengths of their attractors and the number of nodes that
change their states while the network is moving through the attractor.
We consider the networks at the different possible fixed point values
$b^*$. A fixed point of $b$ in the annealed approximation does not
necessarily imply that the network dynamics reaches a fixed point in
state space (i.e. a dynamical fixed point). In our simulations, all
attractors in state space have a constant value of $b$ (with some
fluctuations around it because the system size is finite), which means
that the proportion of nodes changing their state from 1 to 0 is at
each step approximately equal to the proportion of nodes changing
their state from 0 to 1, as suggested by the annealed approximation. 
 We compare the results of the simulations
with the $\lambda$ values calculated within the annealed
approximation.

\subsubsection{$h > 0$}
The only parameter combination for which the considered networks can
be critical is $h = 0.5$ and $K = 2$. The attractors found in the
simulations have lengths of one to four digits, the number of nodes
that change their state on the attractors is of the order of $10^3$,
which is compatible with the expected number of the order of $N^{2/3}$ \cite{kaufmanandco:scaling}.

For $K > 2$ and $h = 0.5$, the networks should be chaotic according to
the annealed approximation. In the simulations, the attractors are
longer than our search range (max.  transient length: $10^5$, max.
attractor length: $10^4$) which is consistent with the expectation of
having a chaotic network.  The annealed approximation predicts
furthermore that networks with larger $h$ are frozen when the only
stable fixed point is $b^* = 0$. Networks with $K$ values for which a
second stable fixed point $b^*>0$ exists have $\lambda > 1$ at this
fixed point and should therefore be chaotic. Simulations of networks
with non-integer $h = 1.5$ and 2.5 show results that are consistent
with this prediction. Quenched networks with integer $h$ show again
deviations.  For $h = 1$ and $K = 12$, even networks with $b$ values
at the second fixed point $b^* > 0$ are frozen and not chaotic.  For
$h = 1$ and $K = 13$, no attractors are found within the search range,
pointing at chaotic dynamics. For $h = 2$ the situation is different.
As stated in the previous section, a second stable fixed point $b^*>0$ appears
already at $K = 31$, and the networks show chaotic behaviour at this
fixed point. According to the annealed approximation, the first
chaotic network should have $K = 35$. The same is true for networks
with $h = 3$: they are chaotic when they reach the second stable fixed
point $b^*$, but this fixed point appears for $K$ values smaller than
predicted by the annealed approximation.

\subsubsection{$h = 0$}
For $h = 0$, the situation is similar to that for $h = 1$. According
to the annealed approximation networks should be chaotic from $K = 13$
on. But in all simulations of networks with connectivities up to $K =
16$ we find only fixed point attractors, which means that these
networks are in the frozen phase.

\subsubsection{$h < 0$}
For $h < 0$, we first chose the values of $h$ and $K$ such that the
networks are expected to be in the frozen phase, and we found fixed
point attractors in all simulations. Then we took a closer look at
parameter values close to the transition between ordered and chaotic
dynamics (cf. figure \ref{phase}). For integer $h$, we found again
deviations from the annealed approximation.  Just as for $h = 0$ and
1, the frozen phase is extended to higher $K$ values. This means that
we find networks with fixed point attractors only in regions of the
parameter space where they should be chaotic according to the annealed
approximation.  For non-integer $h =-0.5$ and $-2.5$, chaotic networks
should be found for $K \ge 5$ and $K \ge 11$ respectively. In all
simulations with these parameters, the attractor lengths exceeded the
search range. For $h = -0.5$ and $K = 4$, on the other side of the
phase boundary where networks are expected to be frozen, we find
mostly short attractors with less than $1\%$ of the nodes changing
their state. The major part of these networks is frozen, in agreement
with the calculated value $\lambda \approx 0.995$. For $h = -2.5$ and
$K = 10$, the situation is similar.  For $h = -1.5$, the annealed
approximation predicts chaotic dynamics for $K \ge 7$.  But since
$\lambda \approx 1.008$ for $K=7$, we are very close to the boundary
for this parameter value. The simulated networks have attractor
lengths ranging from one digit to values exceeding the search
range. For $K = 6$, we have $\lambda \approx 0.903$, and the attractors
are short, with less than $1\%$ of the nodes changing their
values.

To summarize, the phase diagram figure~\ref{phase} obtained by the
annealed approximation is valid for the quenched system for all
non-integer $h$. For integer $h \le 1$, the transition from the frozen to
the chaotic phase occurs at a larger value of $K$ than predicted by the
annealed approximation, and for integer $h>1$ it occurs at a smaller
value. Since most of these transitions do not lie in the window of $K$
values shown in figure~\ref{phase}, 
the corresponding figure obtained from our simulations looks hardly
different, therefore we do not include it.

The reason why the transitions from the frozen to the chaotic phase
do not occur for the $K$ values predicted by the annealed
approximation when $h$ is integer, is the same as the reason why the
stationary values $b_f$ do not agree with the fixed points calculated with the
annealed approximation. For integer $h$, the annealed approximation
does not capture correctly the dynamical properties of the system,
because it neglects memory effects. Even when we evaluate $\lambda$
within the annealed approximation by using the values
$b_f$ obtained from the simulations, the calculated phase transitions
do not occur at the same $K$ values as those obtained by computer
simulations when $h$ is integer.

\section{Discussion}
We investigated the phase diagram of threshold networks with
real-valued thresholds and an updating rule that does not change the
state of a node when the sum of its inputs gives exactly the threshold
value. We compared the analytical results obtained by using the
annealed approximation with the results obtained from computer
simulations. We evaluated the proportion $b$ of 1s in the
networks and the sensitivity $\lambda$ to changes of the state of an
input. We found that the annealed approximation is valid in the case
of non-integer thresholds, but that it does not agree with the
simulations in the case of integer thresholds. We ascribed this
discrepancy to memory effects that are not captured by the annealed
approximation.  In the studies mentioned before \cite{Kuerten88b,
Rohlf02, Rohlf07}, this situation did not occur, and the
annealed approximation was sufficient to calculate the phase
transitions correctly.

Let us now briefly return to the model of the cell-cycle network of
yeast \cite{Li04}, which motivated us to study threshold networks with
this special kind of updating rule. This model consists of 11 nodes,
and it shows seven fixed points. The dominant fixed point corresponds
to the G1 phase of the cell cycle, during which the cell grows.
Although the trajectory corresponding to the cell cycle is
impressively stable, the dominant fixed point is very sensitive to
specific perturbations at certain nodes \cite{Fretter08}. When the
state of one of the 11 nodes is changed, the network returns to the
fixed point only in 6 out of the 11 cases. In the other cases, the
dynamics is attracted to one of the other fixed points of the network.
6 of the 7 fixed points can be reached from other fixed points by
changing only one node, and there is a group of three nodes amongst
which all these changes occur. The inputs of all three nodes are at
the threshold, and changing the state of one of these nodes does not
change the state of any other node. As we have seen in this paper,
such a set of fixed points, which differ by the state of one node, is
characteristic of a threshold network with an integer-valued
threshold. It is due to the update rule that a node keeps its
state when the sum of its inputs is exactly at the threshold. This
raises the question whether the non-dominant fixed points have a
biological meaning, or whether they are just artefacts of the update
rules of the model.

\bigskip
Acknowledgements:
This work was supported by the Deutsche Forschungsgemeinschaft (DFG) under contract no. Dr200/4-1.

\bibliography{bib.bib}
\end{document}